\documentclass[twocolumn,twoside,slac_two]{revtex4}
\usepackage{graphicx}
\usepackage{fancyhdr}
\begin{document}

\title{Reconstruction of High Transverse Momentum Top Quarks at CMS}

%

\author{Gavril Giurgiu}
\affiliation{Department of Physics and Astronomy, Johns Hopkins University, Baltimore, MD, 21218, USA}

\begin{abstract}
High mass resonances decaying into $t \bar{t}$ pairs appear in many extensions of the 
Standard Model. The top quarks from these decays have high transverse momenta and 
their decay products are highly collimated due to the boost into the lab 
frame. As a result the standard techniques for reconstructing $t \bar{t}$ events begin 
to fail. In this talk we discuss the prospects for detecting booted top quarks at CMS.
A new top jet tagging algorithm is presented. 
This algorithm achieves an efficiency of 46\%
for boosted top jets and rejection of 98.5\% for generic QCD jets with transverse 
momenta of 600~GeV/$c$.

\end{abstract}

\maketitle

\thispagestyle{fancy}


\section{Introduction}
Various theoretical extensions of the Standard Model predict the existence of new 
heavy particles which decay into $t \bar{t}$ pairs with large branching fraction. 
Such scenarios include excited neutral gauge bosons $\mathrm{Z'}$ with Standard-Model 
type couplings or Randall-Sundrum KK gluons~\cite{rs}. If these new particles 
are much heavier than the top quark and their masses reach the TeV range, then 
the top quark daughters are highly boosted. 
The jets associated with the boosted top quark decays may be collimated into a single jet. 
In such case, standard methods for identifying top quarks may fail or be severely 
impaired. For instance, b-tagging techniques based on identification of tracks 
or vertexes displaced with respect to the primary interaction vertex would suffer due to the dense 
track environment characteristic to very high energy, collimated jets.
Lower tagging efficiency and higher mistag rates are expected~\cite{btag_high_pt}.
Difficulties in identifying leptons inside boosted jets would diminish the performance 
of lepton based taggers. 

It is therefore very important to develop reconstruction algorithms that 
distinguish boosted top jets from jets produced in generic QCD events. 
We describe an algorithm which attempts to identify boosted top quark jets in which 
the $W$ top daughter decays hadronically. The fraction of such fully hadronic top decays 
is 68\%. 
The idea for tagging boosted top quarks decaying hadronically is to identify jet 
substructure in top quark jets and to use this substructure to impose kinematic cuts 
that discriminate against non-top jets.

\section{Boosted Top Tagging and Cambridge-Aachen Jet Clustering Algorithm}

If a top quark decays fully hadronically $t \rightarrow W^{+} b$ with $W^{+} \rightarrow q \bar{q}'$ 
and the jets from the top quark daughters are collimated into a single top jet, one can try 
to determine the top jet sub-structure by decomposing the top jet into sub-jets corresponding 
to the top daughters $b, q$ and $\bar{q}'$. Once the top jet is decomposed one can attempt to discriminate 
top jets from QCD jets using jet sub-structure information.  

To construct boosted top jets, the Cambridge-Aachen (CA) algorithm~\cite{ca_1} is used. These final CA jets are
referred to as hard jets.
The method developed in Reference~\cite{catop_theory} is implemented to discern the jet sub-structure. 
This approach uses the CA
jet algorithm to reconstruct highly boosted top jets
and decompose them into sub-jets. This decomposition is done by examining the cluster sequence of the final
jets in the CA algorithm to find intermediate sub-jets from
the algorithm, and attempting to identify the jets from the top and $W$ decays. 

The CA algorithm is a $k_{\mathrm{T}}$-like algorithm. These algorithms examine four-vector inputs
pairwise and construct jets hierarchically. To do so, they construct the quantities~\cite{fastjet}:

\begin{eqnarray}
d_{ij} = \mathrm{min}( k_{{\mathrm{T},i}}^n, k_{{\mathrm{T},j}}^n) \frac{\Delta R_{ij}^2}{R^2} \\
d_{i\mathrm{B}} = k_{{\mathrm{T},i}}^n
\end{eqnarray}
where $k_{{\mathrm{T},i}}$ is the transverse momentum of the $i$-th particle with respect
to the beam axis,
$\Delta R_{ij}$ is the distance between
particles $i$ and $j$ in $(y,\phi)$ space (where $y$ is rapidity,
and $\phi$ is the azimuthal angle), and $R$ is a distance parameter
taken of order unity. 
For the $k_{\mathrm{T}}$ algorithm, $n = 2$. 
For the anti-$k_{\mathrm{T}}$ algorithm, $n=-2$. 
For the CA algorithm, $n=0$ and $d_{i\mathrm{B}} = 1$.
The quantity $d_{i\mathrm{B}}$ is referred
to as the beam distance. 
The algorithm then finds the minimum $d_{\mathrm{min}}$ of all the $d_{ij}$ and $d_{i\mathrm{B}}$. 
If $d_{\mathrm{min}}$ is a $d_{ij}$, the two particles are merged (by default, via a four-vector
summation). If it is a $d_{i\mathrm{B}}$, then the particle $i$ is a final jet, and is removed
from the list. This process is repeated until there are no particles left. 
In the case of a CA algorithm with $R = 0.8$, the merging condition ($d_{ij} < d_{iB}$) 
reduces to $\Delta R < 0.8$.

The final hard jets are required to have transverse momentum above 250 GeV/c  
and rapidity within the $\pm2.5$ range. The sub-jets are selected if the sub-jet transverse 
momentum is larger than 0.05 the hard jet transverse momentum, 
$P_T$(sub-jet)$ > 0.05 \times P_T$(hard\_jet). The top tagging algorithm is 
applied if at least three sub-jets are found.   

The variables that are used to discriminate top jets from generic QCD jets are: 
the number of sub-jets identified inside the hard jet, the hard jet mass (as proxy to the top mass) 
and the minimum di-jet mass pair among the three leading sub-jets (as proxy to W mass).

\begin{figure}[h]
\centering
\includegraphics[width=80mm]{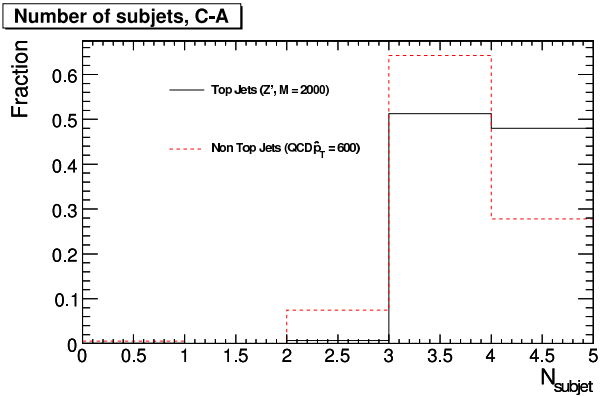}
\caption{ Number of sub-jets inside boosted top jets from 2~TeV $Z'$ decays $Z' \rightarrow t \bar{t}$ (black, solid line) 
	versus non-top jets from generic QCD (red, dashed line).
	The samples are chosen such that the reconstructed top and QCD jets have approximately 
	the same transverse momenta. } 
\label{n_subjets}
\end{figure}

Figure~\ref{n_subjets} shows the distribution of the number of sub-jets in collimated top 
jets from a 2~TeV/c${^2}$ mass $Z'$ resonance compared to the corresponding distribution of 
generic QCD jets. The QCD jets are selected so that their transverse momenta are in the 
same range as the typical transverse momenta of the top quarks from the $Z'$ resonance. 
A requirement that the hard jet contains at least three 
sub-jets is applied as it rejects a significant fraction of QCD background 
jets and retains most of the top jet signal events. 

\begin{figure}[h]
\centering
\includegraphics[width=80mm]{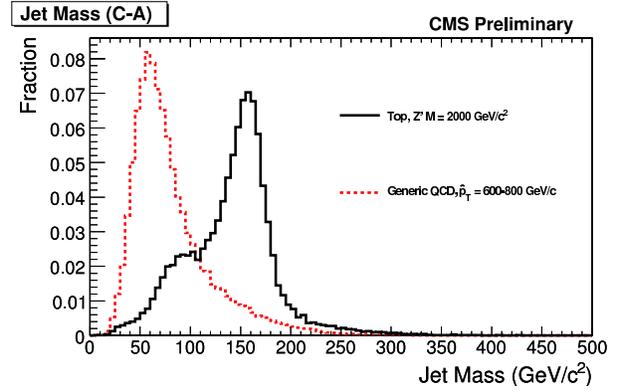}
\caption{ Jet mass distributions for boosted top jets from 2~TeV $Z'$ decaying as 
	$Z' \rightarrow t \bar{t}$ (black, solid line) and generic QCD jets 
	(red, dashed line).} 
\label{jet_mass}
\end{figure}

The use of the jet mass as discriminating variable between top and QCD jets is justified because, in the case of
true top jets, the jet mass tends toward the top mass, while
for generic QCD non-top jets, 
the jet mass does not reconstruct to the top mass but instead
approximately scales by the jet transverse momentum over a constant 
of order 10. Figure~\ref{jet_mass} shows the distributions of the hard jet 
mass for top jets from a 2~TeV/c$^2$ $Z'$ resonance and the corresponding distribution 
for QCD jets with transverse momenta similar to the top jets. Hard jets with masses between   
100 and 250 GeV/c are selected.   

\begin{figure}[h]
\centering
\includegraphics[width=80mm]{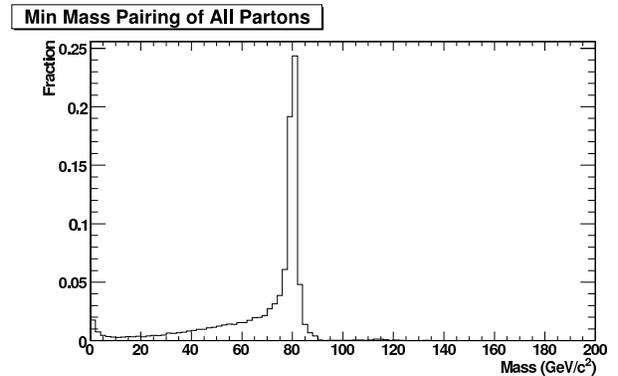}
\caption{Distribution of the minimum di-jet invariant mass. The $W$ mass is reconstructed 
	 in most cases. } 
\label{gen_min_mass}
\end{figure}

The minimum pairwise mass of the sub-jets often reconstructs in the vicinity of the $W$ mass.
Figure~\ref{gen_min_mass} shows the true minimum mass pairing
of the three partons from the $t \rightarrow Wb \rightarrow q\bar{q}'b$ decay
where the top quarks come from the ${\rm Z'}$ sample. It is most often
the case that the minimum mass pairing of the true partons
results in the $W$ mass, which means that the $b$ quark is most often
the hardest parton in the event. Despite the fact that the lowest mass pairing
of the sub-jets is not always the $W$ mass after hadronization and reconstruction,
the minimum mass pairing selection criterion is nonetheless
exploited. The minimum mass pairing provides good discrimination
against non-top jets, where there is no on-shell $W$ and 
instead the minimum mass pairing of the sub-jets
reconstructs to a low-mass falling spectrum.

\begin{figure}[h]
\centering
\includegraphics[width=80mm]{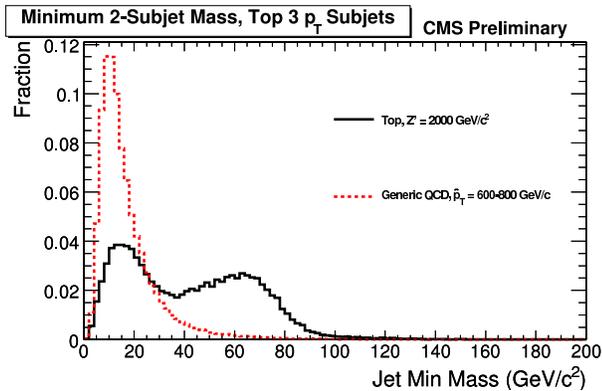}
\caption{ Distributions of the minimum di-jet invariant mass from boosted top quarks (black 
	solid line) and from QCD jets (red, dashed line). Among the sub-jets inside the hard jet, 
	the three with the highest transverse momenta are selected. The invariant mass of each 
	pair of two sub-jets is calculated. Among the three sub-jet pairs, the one with minimum 
	di-jet mass is chosen as proxy to the $W$ mass. In the case of top jets, besides a low mass 
	peak at $\approx 10$~GeV/c, a clear peak from $W$ decays 
	is also seen below the W mass at $\approx 65$~GeV/c. In the case of QCD jets only the 
	low mass peak at $\approx 10$~GeV/c is observed.  
	  } 
\label{compare_jet_min_mass}
\end{figure}

Figure~\ref{compare_jet_min_mass} shows the minimum pairwise mass of the three 
reconstructed sub-jets with the highest transverse momenta for 
top jets from $Z' \rightarrow t \bar{t}$ decays versus non-top jets 
from generic QCD samples, respectively. 
The minimum pairwise mass is required to be above 50 GeV/c$^2$.
This minimum di-jet mass requirement is chosen to optimize $S/\sqrt{B}$ where 
$S$ is the number of top jets and $B$ is the number of background QCD events.

\section{Boosted Top Tagging Performance}

\subsection{Efficiency}

\begin{figure}[h]
\centering
\includegraphics[width=80mm]{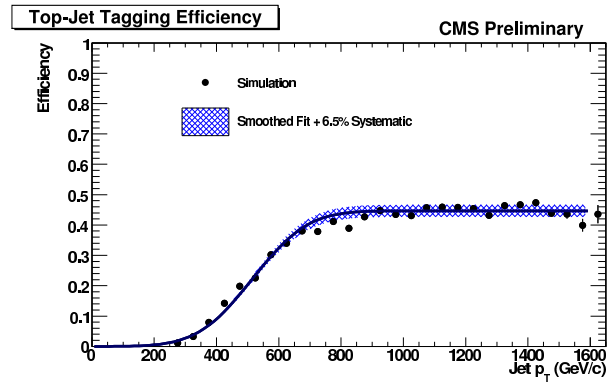}
\caption{Top tagging efficiency as function of the top jet transverse momentum.}
\label{fig:efficiency}\end{figure}

To estimate the efficiency of the boosted top tagging algorithm 
several simulated samples of Randall-Sundrum gluons 
decaying to $t \bar{t}$, with masses in the range 750-3000 GeV/c$^2$ were examined. 
The efficiency defined as the number of matched top-jets
that are identified by the algorithm divided by the total number of matched
top-jets is measured on these samples as function of jet transverse momentum. 
The efficiency as function of the top jet transverse momentum 
is shown in Figure~\ref{fig:efficiency}. The efficiency reaches a plateau value 
of $\approx45\%$ for jet transverse momenta above $\approx700$~GeV/c. Below 600-700 GeV/c the 
efficiency is lower and drops to zero below 300 GeV/c. This behavior is explained 
by the fact that this algorithm requires the daughters of the boosted top quark 
to be merged into a single jet. Merging is enhanced as the top quark momentum 
increases. For low transverse 
momenta the top quark daughters produce separate jets. 
As their transverse momenta increase from $\approx300$~GeV/c to $\approx700$~GeV/c the 
top jets become more and more collimated, approaching full merging above $\approx700$~GeV/c.  

\subsubsection{Theoretical Systematic Uncertainties}

There are several theoretical systematic effects that can affect the estimate of the top 
tagging efficiency by changing the profile of the sub-jets: 

\begin{itemize}
\item Initial and final state radiation
\item Renormalization scale 
\item Fragmentation
\end{itemize}

The issue is that considering a reasonable variation on these parameters is not
yet understood. Variations are taken relying on experience from lower
energy colliders extended with theoretical arguments. 
A total theoretical uncertainty of 3.8\% is found.
This estimate should be taken only as indicative of the
theoretical uncertainty, while a more careful study must be determined
in the future to ascertain a more accurate estimate, when there is
sufficient data to estimate these effects. 

\subsubsection{Detector-based Systematic Uncertainty} 

In order to account for the detector-based systematic uncertainties, the resolution
of the sub-jets within the hard jets was derived from simulation of
$Z'\rightarrow t \bar{t}$ events with masses of 1000 and 3000 GeV/$c^2$. 
The partons
from the $t\rightarrow W b\rightarrow b q \bar{q}'$ decay (i.e. the $b$, $q$, and $\bar{q}'$)
were matched to the closest reconstructed sub-jet. The response of the
simulated calorimeter was then parameterized with sub-jet transverse momentum. This was
done for the resolution of the transverse momentum, rapidity, and azimuthal angle. 

It was observed that the resolutions could be estimated as
\begin{eqnarray}
\frac{\sigma}{p_T} = \frac{74\%}{\sqrt{p_T - 24}} \oplus 15\%, \\
\sigma(y) = \frac{41\%}{\sqrt{p_T - 25}} \;\oplus\; 1.3\%\; \oplus\; 6.5\times 10^{-5} p_T, \\
\sigma(\phi) = \frac{44\%}{\sqrt{p_T - 25}} \;\oplus\; 0.0\%\; \oplus\; 5.6\times 10^{-5} p_T. 
\end{eqnarray}

The resolutions were hypothesized to be 10\% and 50\% worse than the simulation
for the momentum and angular resolution, respectively. 
An additional 5.3\% systematic uncertainty due to assumed worse resolution was assigned 
to the efficiency. 

\subsubsection{Total Systematic Uncertainty}

Figure~\ref{fig:efficiency} shows the efficiency
with simulation statistical uncertainties, as well as the total 6.5\% systematic
uncertainty from combining the theoretical (3.8\%) and detector-based (5.3\%) 
systematic uncertainties. Table~\ref{tab:eff_systematics} summarizes the
systematic uncertainties. 

\begin{table}[h]
\caption{\label{tab:eff_systematics} Effects of variation of several systematic
	uncertainties
	on the estimated efficiency from simulation.}
    \begin{center}
    \begin{tabular}{|l|c|} 
        \hline
        Effect                       & Systematic Uncertainty (\%) \\
        \hline
        Initial State Radiation      & 1 \\
        Final State Radiation        & 2 \\
        Renormalization Scale        & 3 \\
        Light Quark Fragmentation    & $< 1$ \\
        Heavy Quark Fragmentation    & $< 1$ \\ 
        \hline
	Theoretical Uncertainty      & 3.8 \\
        \hline
	Momentum Smearing + 10\%     & 3.3  \\
	Azimuthal Smearing + 50\%    & 2.9  \\
	Rapidity Smearing + 50\%     & 2.9  \\
        \hline
        Detector-Based Uncertainty   & 5.3 \\
	\hline
	Total Systematic Uncertainty & 6.5 \\
	\hline
    \end{tabular}
    \end{center}

\end{table}

\subsubsection{Efficiency Cross Checks}

The shape of the efficiency curve has been studied, and the primary factor 
has been determined to be the R-parameter in the CA algorithm.
As the width of the jet is increased, the lower transverse momentum top jets
have more products merged. However, at some point at higher transverse momentum
values, the only quantities that are subsumed by a larger
distance parameter are radiative jets, which manifests in a decreasing
efficiency because the minimum mass combination of the sub-jets
tends to bias away from the $W$ mass when there is radiation present.

\begin{figure}[h]
\centering
\includegraphics[width=80mm]{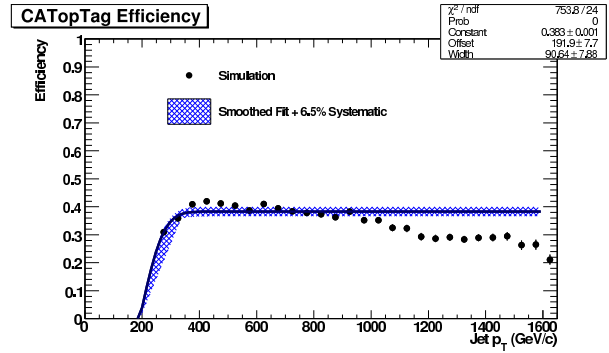}
\caption{ The efficiency turn-on for a distance parameter R = 1.5 (up from the default 0.8).
	The faster turn-on at the low transverse momentum end is readily apparent,
	as is the faster turnoff at the high transverse momentum end.} 
\label{fig:eff_catoptag_widecone}
\end{figure}

Figure~\ref{fig:eff_catoptag_widecone} shows the efficiency turn-on
for a distance parameter of 1.5 (up from the default 0.8).
The faster turn-on at the low transverse momentum end is readily apparent,
as is the faster turn-off at the high transverse momentum end. 

\subsection{Fake Tag Rate}

Non-top decays may pass the selection defined in the previous section and thus fake a 
boosted top tag. 
In order to derive a parameterization of the fake tag rate, a data-driven method is proposed
that makes use of a high statistics sample, and uses an ``anti-tag and probe'' method.
This method is expected to provide over a thousand fake tags for a data sample of  100~pb$^{-1}$, 
allowing for  a robust  data driven determination of the fake background.

The following selection is made to select fake tags:
\begin{itemize}
\item Two jets are required to have $p_T >$ 250 GeV/c, and $|y| < 2.5$.
\item Events are required to have one jet ``anti-tagged''. 
      To ``anti-tag'',  jets are selected that have two sub-jets or less,
      or to have more than two sub-jets, with jet mass and jet minimum mass outside the signal window.
\item The other jets in the sample are referred to as the ``probe'' jets. The
      contamination from continuum $t \bar{t}$ production is 
      subtracted based on an estimate from simulation, and the amount of that subtraction 
      is taken as a systematic uncertainty. 
      This ``probe jet'' selection constitutes an almost entirely signal-depleted sample.
\item The tag rates are then parameterized with respect to the jet $p_T$ using these ``probe jets''. 
The prediction from the simulation is taken as the central value and scaled to 100 pb$^{-1}$,
assuming Poisson statistics and taking a binomial uncertainty. 
\end{itemize}

\begin{figure}[h]
\centering
\includegraphics[width=80mm]{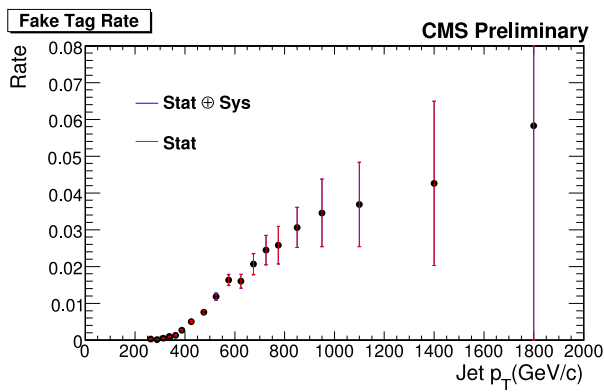}
\caption{ The fake tag parameterization as function of transverse momentum for a 
	100~pb$^{-1}$ data sample.} 
\label{fig:mistag_param_lum}
\end{figure}

Figure~\ref{fig:mistag_param_lum} shows the fake tag 
parameterization as function of transverse momentum for a 100 pb$^{-1}$ data sample. These plots 
should be taken as a proxy for the real data. The results are fully data-driven in 
the real analysis with data, with the sole exception of the correction for 
the $t \bar{t}$ contamination. 
Even for a sample as low as  100 pb$^{-1}$, it is possible to reliably estimate the
fake tag rate directly from the data, with an approximately 33\% statistical
uncertainty for jets with $p_T = 800$ GeV/$c$.

\section{Conclusions}

The algorithm described in Ref.~\cite{catop_theory} has been implemented in CMS
and has achieved similar rejection of non-top backgrounds as described in that
paper. 

The algorithm deals exclusively with hadronic decays of the $W$ boson in the
cascade decays of top quarks, and has made this channel accessible experimentally, due to its
high rejection ($\approx$ 98\% of jets with $p_T = 600$ GeV/$c$) 
of non-top-quark boosted jets while retaining a high fraction of top-quark boosted
jets ($\approx$ 46\% of jets with $p_T > 600$ GeV/$c$). 
This performance is comparable to that for bottom-quark jet-tagging algorithms at
hadron colliders.


\bigskip 

\end{document}